\documentclass[reprint,aps,prd,nofootinbib,a4paper,10pt,twocolumn, superscriptaddress,floatfix]{revtex4-2}
\usepackage{hyperref} 
\usepackage{changes}
\usepackage{graphicx}
\usepackage{caption}
\usepackage{subcaption}
\usepackage[section]{placeins}
\usepackage[titletoc]{appendix}
\usepackage{xcolor}
\usepackage{dsfont}
\usepackage{bm}
\usepackage{mathtools} 
\usepackage{amsfonts}
\usepackage{enumerate}
\hypersetup{
     colorlinks   = true,
     citecolor    = cyan,
     linkcolor = pink
     }
\begin{document} 

\title{Safely simplifying redshift drift computations in inhomogeneous cosmologies: Insights from LTB Swiss-cheese models}

\author{David Rønne Sallingboe}
\email{sallingboe.david@gmail.com}
\author{Sofie Marie Koksbang}
\email{koksbang@cp3.sdu.dk}
\affiliation{Section for Fundamental Physics, Department of Physics, Chemistry and Pharmacy, University of Southern Denmark, Campusvej 55, DK-5230 Odense M, Denmark}

\begin{abstract}
One of the most important discoveries in cosmology is the accelerated expansion of the Universe. Yet, the accelerated expansion has only ever been measured {\em in}directly. Redshift drift offers a direct observational probe of the Universe's expansion history, with its sign revealing whether there has been acceleration or deceleration between source and observer. Its detection will mark a major milestone in cosmology, offering the first direct measurement of an evolving expansion rate. Given its epistemic importance, it is essential to understand how its measurements can be biased. One possibility of bias comes from cosmological structures. However, theoretical estimates of such effects are difficult to obtain because computing redshift drift in general inhomogeneous cosmologies is computationally demanding, requiring the solution of 24 coupled ordinary differential equations. In this work, we use Lemaitre-Tolman-Bondi Swiss-cheese models to show that only the two dominant contributions are needed to achieve percent-level accuracy up to $z = 1$. This allows the reduction of the full system of 24 ODEs to the standard null geodesic equations, significantly simplifying the calculation.
\newline\indent
Although our analysis is based on idealized Swiss-cheese models with spherical structures, we expect that similar simplifications apply to more complex scenarios, including cosmological N-body simulations. Our analysis thus underpins a practical and robust framework for efficient redshift drift computations applicable to a wider range of inhomogeneous cosmologies.
\end{abstract}

\maketitle

\section{Introduction}
One of the most profound discoveries in cosmology is that the Universe is expanding. Perhaps even more remarkably, in 1998, observations of distant supernovae revealed that this expansion is accelerating \cite{sn1,sn2}, contradicting expectations at the time. However, to date, all evidence for cosmic acceleration remains indirect. For example, the original discovery relied on supernovae appearing slightly dimmer and hence more distant than expected in a decelerating model such as the Einstein-de Sitter model. More recently, tensions have emerged between different cosmological observations when interpreted within the standard model \cite{challenges}, and some studies even question whether current data truly supports accelerated expansion \cite{no_acc1, no_acc2, no_acc3, no_acc4, no_acc5}.
\newline\indent
Given the deep implications of cosmic acceleration for spacetime, gravity, and fundamental physics, a direct measurement is essential. This is the promise of real-time cosmology, a transformative approach that will, for the first time, allow us to observe the Universe’s expansion {\em as it happens} \cite{realtime}. Unlike traditional methods that infer cosmic dynamics e.g. through relations integrating over the expansion rate, real-time cosmology seeks to measure the evolution directly. A key observable in this context is redshift drift: the gradual change in the redshift of distant sources due to the evolving expansion rate of the Universe. Because this change is extremely slow, its detection requires exceptionally precise instruments. When redshift drift was first proposed in 1962 \cite{sandage, Mcvittie}, estimates suggested that millions of years of observation time would be needed. Today, thanks to advances in observational technology, facilities such as the SKA and ELT are expected to detect this signal within the coming decades \cite{SKA, ANDES, ANDES2, ELT, SKAogELT, SKAnew}. Other instruments, such as CHIME and FAST, may be upgraded to contribute \cite{fast, chime}, and some even advocate for dedicated missions given the foundational importance of this observable \cite{whitepaper}.
\newline\indent
While redshift drift can help constrain cosmological parameters \cite{SKAogELT}, its core value lies in providing a direct measurement of how the expansion rate evolves over time. In a homogeneous and isotropic Friedmann–Lemaitre–Robertson–Walker (FLRW) spacetime, the redshift drift, $\delta z$, is given by 
\begin{align}\label{eq:dz}
\frac{\delta z}{\delta t_0} = (1+z) \left(\dot a|_0 - \dot a|_e \right),
\end{align}
where subscripts $e$ and $0$ denote evaluation at the emitter and observer, respectively. Here, $a$ is the scale factor, $z$ is the redshift measured at present time $t_0$, and $\delta t_0$ is the interval between the two redshift measurements. The sign of the redshift drift directly reflects the change in the expansion rate: a positive sign indicates acceleration between emission and observation, while a negative sign indicates deceleration. A non-positive redshift drift at low redshifts would severely challenge the standard model of cosmology and lend support to alternative scenarios \cite{melia, anotherlook}.
\newline\indent
Regardless of its sign, the detection of redshift drift would represent the first direct observation of evolving cosmic expansion and thus be a major breakthrough in cosmology. However, equation \ref{eq:dz} assumes a perfectly homogeneous and isotropic Universe, neglecting the influence of cosmic structures such as galaxies and clusters. In reality, inhomogeneities induce fluctuations in the redshift drift, potentially biasing measurements away from the idealized FLRW prediction. To accurately interpret future observations, it is therefore crucial to quantify the impact of structures on redshift drift measurements. A major obstacle for this is that the exact computation of redshift drift in a general inhomogeneous cosmological spacetime requires solving 24 coupled ordinary differential equations (ODEs) plus four extra ODEs to track source peculiar velocity \cite{selv_dzLTB}. Significant progress has been made towards simplifying this calculation. In \cite{selv_dzLTB} is was shown that for observers located in the homogeneous FLRW background outside a Lemaitre-Tolman-Bondi (LTB) \cite{ltb1, ltb2, ltb3} double-structure, the most computationally intensive terms could be safely neglected. However, a follow-up study in \cite{selv_gadget}, based on a limited set of light rays, demonstrated that these terms become more significant when the observer is situated within the inhomogeneous region. This is an important insight, since real observers reside within structures rather than in idealized backgrounds.
\newline\indent
Building on this foundation, the work presented here offers a comprehensive analysis of redshift drift in an LTB Swiss-cheese model. We systematically assess the relative importance of each contribution to the redshift drift by simulating light propagation through a Swiss-cheese model with LTB structures, placing observers both inside and outside the inhomogeneities. Our results provide a critical step toward identifying which contributions can be safely neglected over longer redshift intervals, thereby enabling significant simplification of redshift drift computations in realistic cosmological settings.
\newline\newline
In section \ref{sec:model_setup}, we review the LTB models and describe the Swiss-cheese setup we use for our analysis. Section \ref{sec:light} presents the theoretical framework for computing redshift drift in the Swiss-cheese model. Section \ref{sec:results} contains our numerical findings, and section \ref{sec:summary} concludes with a summary and discussion.

\section{Model setup}\label{sec:model_setup}
In this section, we specify the LTB structure and Swiss-cheese configuration used in our study.
\newline\newline
The LTB models represent spherically symmetric (comoving) dust solutions to Einstein's field equation. Using spherical coordinates centered on the symmetry center of the LTB model, the line element of the LTB spacetime can be written as 
\begin{align}
    ds^2 = -c^2dt^2 + \frac{R_{,r}^2}{1-k(r)}dr^2 + R^2 d\Omega^2.
\end{align}
Following \cite{book}, we can write the corresponding solutions to Einstein's field equation as
\begin{equation}\label{eq:R_of_etha}
\begin{aligned}
k&>0:\\
&R = \frac{M}{k}\left( 1-\cos\left( \eta\right) \right) ,\, \,  \eta -\sin\left( \eta\right) =\frac{ck^{3/2}}{M}\left( t-t_{B}\right) \\
k&=0:\\
&R = \left(\frac{9}{2}Mc^2\left( t-t_{B}\right)^2  \right) ^{1/3}\\
k&<0:\\
&R = \frac{M}{-k}\left(\cosh\left( \eta\right) -1 \right)  ,\, \,  \sinh\left( \eta\right) -\eta = \frac{c\left( -k\right) ^{3/2}}{M}\left( t-t_{B}\right). 
\end{aligned}
\end{equation}
We will specify a single LTB model by setting its big bang time ($t_B$) equal to zero. By setting $t_B$ to a constant value, we ensure the Big Bang happens everywhere at once in the model. The value of $t_B$ is for simplicity set to 0 so that coordinate time represents cosmic time. We will furthermore set
\begin{align}
    M = \frac{1}{2}\frac{H_0^2r^3}{c^2}
\end{align}
and
\begin{align}
k(r) = \left\{ \begin{array}{rl}
-k_{\rm max}r^2\left(\left(\frac{r}{r_b} \right)^4 -1 \right)^4  &\text{if} \,\, r<r_b \\
0 &\mbox{ otherwise}.
\end{array} \right.
\end{align}
When specifying $M$ we use $H_0 = 70$km/s/Mpc. To fully specify the model we choose $r_b = 48$Mpc and $k_{\rm max} = 1.3\cdot 10^{-7}$. The resulting density profile of the spacetime is obtained as (see \cite{book})
\begin{align}
    \rho(r) = \frac{c^4}{4\pi G}\frac{M_{,r}}{R^2R_{,r}},
\end{align}
where partial derivatives are denoted by a comma followed by the coordinate. The resulting LTB model represents a double structure of a central void/underdensity surrounded by an overdensity. Outside the overdensity, the model reduces exactly to an Einstein-de Sitter (EdS) model with $H_0 = 70$km/s/Mps. The central void has a present-time depth of approximately $\rho/\rho_{\rm EdS} \approx 0.25$ and a present-time proper radius slightly below $48$Mpc. Our Swiss-cheese model presented below thus has a homogeneity scale of $\sim 100$Mpc. The void is surrounded by an overdensity with a peak density contrast of approximately $\rho/\rho_{\rm EdS} \approx 8$.

\begin{figure}[tb]
    \centering
    \includegraphics[width=0.5\textwidth]{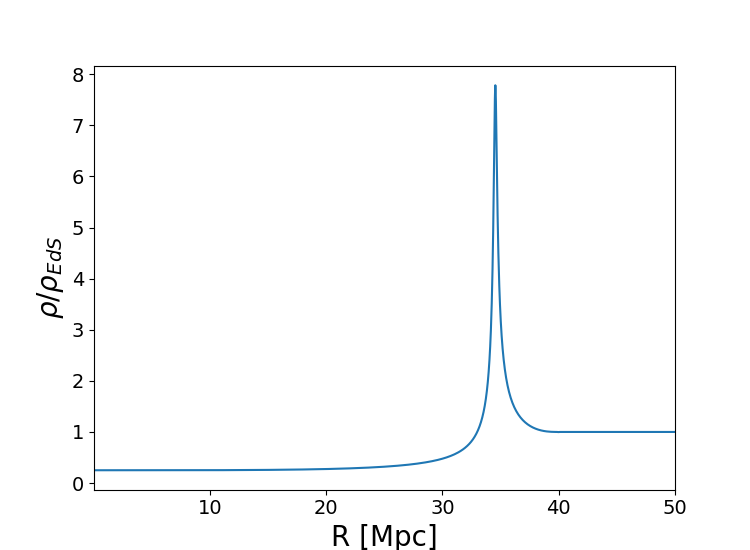}
    \caption{Present-time density profile of the considered LTB model. The horizontal axis shows $R$, which to a good approximation corresponds to the proper radial distance.}
    \label{fig:dens_profile}
\end{figure}

\subsection{Swiss-cheese model}
The LTB model specified above represents only a single double-structure of an underdensity surrounded by an overdensity. To propagate light rays through multiple structures, we can build a Swiss-cheese model where copies of the LTB structure are distributed in an Einstein-de Sitter (EdS) background. We follow \cite{selv_packing} to obtain a high packing fraction and distribute the LTB structures through a random close-packing procedure using the Jodrey-Tory algorithm \cite{torey1, torey2}. A detailed description of the procedure is given in appendix \ref{app:packing}. The final packing fraction of our Swiss-cheese model is slightly above 40 \%. This packing fraction is obtained by distributing 2000 copies of the LTB double-structure in a periodic box of present time side length $1367$ Mpc.

\section{Light propagation}\label{sec:light}
Equation \ref{eq:dz} for the redshift drift is derived under the assumption of an FLRW universe and is not valid in more general models that include structures such as our Swiss-cheese model. In this section, we present the method used for computing the redshift drift in our Swiss-cheese model.
\newline\newline
The first step is to follow the trajectories of light rays as they traverse the Swiss-cheese spacetime. This is achieved by solving the null-geodesic equations
\begin{align}
\frac{dk_{\mu}}{d\lambda} = \frac{1}{2} g_{\alpha\beta,\mu}k^\alpha k^\beta. 
\end{align}
As first noted in \cite{selv_hellaby} (see also the discussion in \cite{selv_dzLTB}), for comoving sources and observers in a spacetime with $g_{tt} = -c^2$ and $g_{ti} = 0$, we can directly calculate the redshift drift as
\begin{align}\label{eq:dz_direct}
    \delta z = \frac{\delta t_0}{k^t_{\rm 0}}\left( -(1+z)(k^t_{t})_0+ \frac{(k^t_{,t})|_e}{1+z}    \right).
\end{align}
To obtain the partial derivative $k^t_{,t}$ we need to solve a set of 16 coupled ODEs simultaneously with the geodesic equations. These 16 additional ODEs are given by \cite{ishak}
\begin{align}
    \frac{dk^{\mu}_{,\nu}}{d\lambda} = \frac{\partial}{\partial x^{\nu}}\frac{dk^\mu}{d\lambda} -k^\beta_{,\nu}k^{\mu}_{,\beta}.
\end{align}
An alternative calculation of the redshift drift was presented in \cite{Asta_cosmography}. There, the redshift drift was written as the integral (setting c = 1 for simplicity)
\begin{align}\label{eq:integral_dz}
\begin{split}
\delta z &= \delta \tau_0 E_e \int_{\lambda_e}^{\lambda_0} d \lambda \big(  -\kappa^{\mu} \kappa_{\mu} +\Sigma^{\bf O}    +  e^\mu \Sigma^{\bf e}_\mu   \\ &+       e^\mu   e^\nu \Sigma^{\bf ee}_{\mu \nu} + e^{\mu}\kappa^{\nu}\Sigma^{\bf e\kappa}_{\mu\nu} \big) ,
\end{split}
\end{align}
where
\begin{align}\label{eq:components}
&\Sigma^{\bf O} :=  - \frac{1}{3} u^\mu u^\nu R_{\mu \nu}     + \frac{1}{3}D_{\mu} a^{\mu} + \frac{1}{3} a^\mu a_\mu    \\   
&\Sigma^{\bf e}_\mu  :=  - \frac{1}{3}  \theta a_\mu    -   a^{ \nu} \sigma_{\mu \nu}  + 3 a^{ \nu} \omega_{\mu \nu}    - h^{\nu}_{\mu} \dot{a}_\nu     \\   
&\Sigma^{\bf ee}_{\mu \nu} :=     a_{ \mu}a_{\nu }  + D_{  \mu} a_{\nu }    -  u^\rho u^\sigma  C_{\rho \mu \sigma \nu}   -  \frac{1}{2} h^{\alpha}_{\,\mu} h^{\beta}_{\, \nu}  R_{ \alpha \beta }   \\
& \Sigma^{\bf e\kappa}_ {\mu\nu} :=2(\sigma_{\mu\nu}-\omega_{\mu\nu})\\
&\kappa^{\mu} = h^{\mu}_{\nu}\dot e^{\nu}.   
\end{align}
The various quantities above are defined in the usual manner such that $R_{\mu\nu}$ denotes the Ricci tensor, and $C_{\rho\mu\sigma\nu}$ the Weyl tensor. The spatial covariant derivative is denoted by $D_{\mu}$ while $h_{\mu\nu}$ is a tensor projecting onto spatial hypersurfaces orthogonal to the velocity field $u^{\mu}$. The acceleration is $a^{\mu}$. The fluid's expansion tensor is $\theta$ and $\sigma_{\mu\nu}$ its shear, while $\omega_{\mu\nu}$ denotes the vorticity. The 4-vector $e^{\mu}$ is the spatial direction vector of the light ray as seen by the observer with velocity $u^{\mu}$.
\newline\indent
Most of the above terms vanish identically in LTB models so that we are left with
\begin{align}\label{eq:components_simplified}
\Sigma^{\bf O} &= - \frac{1}{3} u^\mu u^\nu R_{\mu \nu} \\   
\Sigma^{\bf e}_\mu  &= 0 \\   
\Sigma^{\bf ee}_{\mu \nu} &= -  u^\rho u^\sigma  C_{\rho \mu \sigma \nu}  \\
\Sigma^{\bf e\kappa}_{\mu\nu} &=2 \sigma_{\mu\nu} \\
   \kappa^{\mu} &= - \frac{c}{E} \left( \delta_{\mu \nu} - \frac{c^2 k^\mu k_\nu}{E^2} + \frac{k^\mu u_\nu + u^\mu k_\nu}{E}  k^\nu_{,t} + \Gamma^\nu_{t \beta} k^\beta \right) .
\end{align}
In order to compute the position drift, $\kappa^\mu$, we again need $k^\mu_{,\nu}$. Thus, regardless of which of the two methods is used for calculating the redshift drift, we need to solve the set of 24 coupled ODEs presented above\footnote{In principle, we need four more ODEs to track the 4-velocity of the emitter. However, as demonstrated in \cite{selv_dzLTB, selv_gadget}, we may safely assume that the emitter velocity is that of the dust for LTB models with the initial conditions we use for light propagation here. We therefore omit these additional four ODEs here.}. However, we aim to justify that the redshift drift to high precision can be computed using a simplified version of equation \ref{eq:integral_dz}, where terms involving partial derivatives of the tangent vector are neglected. In the LTB limit, this amounts to making the approximation
\begin{align}
    \delta z\approx -\delta \tau_0 E_e \int_{\lambda_e}^{\lambda_0} d \lambda \left(  \frac{1}{3}u^\mu u^\nu R_{\mu\nu}  +    e^\mu e^\nu u^\rho u^\sigma C_{\rho\mu\sigma\nu}   \right).
\end{align}
In the following, we will thus compute the redshift drift using equation \ref{eq:integral_dz} and compare the sizes of the individual terms. When analyzing our results, we also compared with the direct redshift drift expression in equation \ref{eq:dz_direct} as a consistency check.
\newline\newline
To solve the 24 ODEs, we first need to choose sensible initial conditions. We consider two separate sets of initial conditions: One set for observers in the cheese (EdS background) and one for observers in the holes/LTB structures. The initial conditions are specified below.

\subsubsection{Initial conditions for observer in the cheese}
When the observer is positioned in the cheese, we use the initial condition convention that $k^t_{\rm ic} = -1/c$. We use Cartesian coordinates in the cheese and use random initial conditions for $k^x , k^y, k^z$, normalizing these to ensure that $k^\mu k_\mu = 0$. Following \cite{selv_dzLTB, selv_hellaby}, we then set $k^\mu_{,i} = 0$ and $k^\mu_{,t} = \frac{1}{k^t} \frac{dk^\mu}{d\lambda}$ initially. We use random spatial positions for the observers, which are all placed at present, $t_0$.

\subsubsection{Initial conditions for observer in the holes}
When the observer is positioned in the LTB structures, we will always use the initial condition convention that $k^t_{\rm ic} = -1/c$, just as observers in the cheese. The remaining initial conditions become more complicated when the observer is in an LTB structure, since the initial conditions specified above for the observer in the cheese would break the constraints of the system of ODEs such as the null-condition and its partial derivatives.
\newline\indent
\begin{figure*}[tb]
    \centering
    \includegraphics[width=0.48\textwidth]{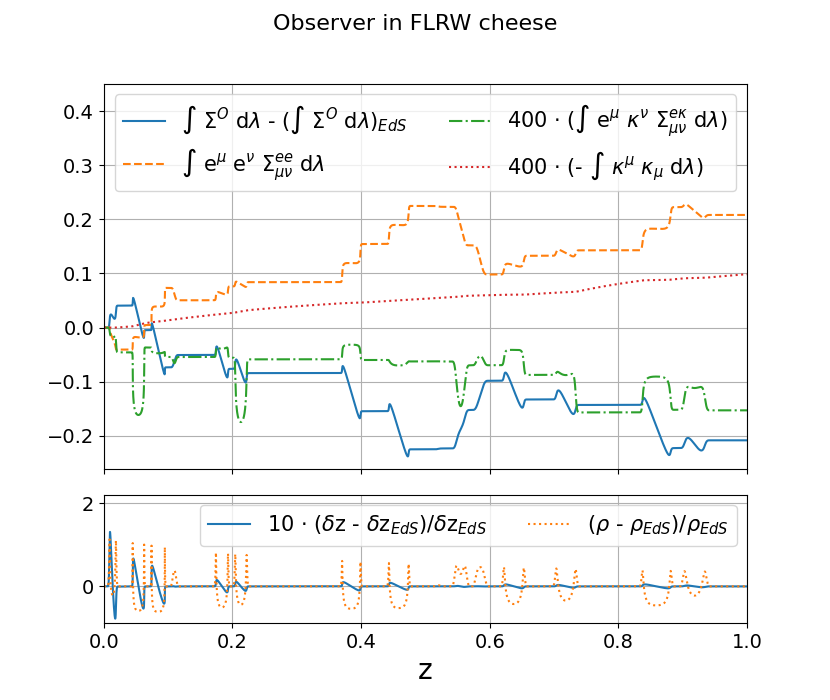}
        \includegraphics[width=0.48\textwidth]{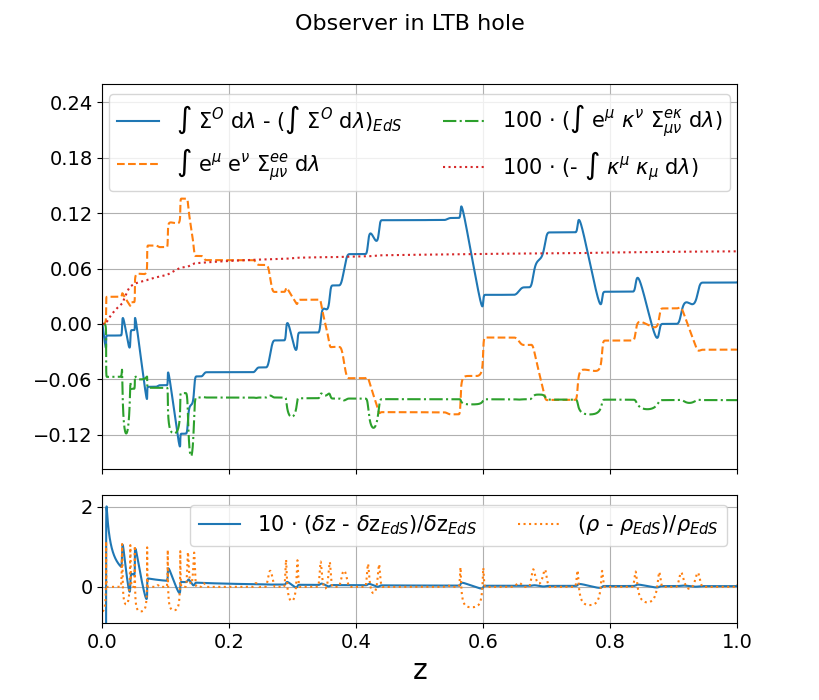}
    \caption{Individual terms in the integral expression for the redshift drift (equation \ref{eq:integral_dz}) for fiducial light rays with observer placed in the cheese and hole. The bottom panels show the total redshift drift relative to the EdS redshift drift, together with the density fluctuations along the light rays.}
    \label{fig:single_ray}
\end{figure*}
For an observer in an LTB structure, we will, for simplicity, set $k^\phi = 0$ initially. For a single LTB model, this does not lead to loss of generality since the LTB models are spherically symmetric. For the Swiss-cheese model, we lose some global generality this way, but since i) we are interested in statistical results from ensembles of light rays and not sky-maps, and ii) the structures in our Swiss-cheese model and the observer positions are random, this does not have any qualitative impact on our results. We now choose $k^\theta = 1/r\cdot \cos(\alpha)$, where $\alpha$ is sampled according to $\cos(\alpha) = 2u-1$, and $u\in[0,1]$ is a new random number for each light ray. We then obtain $k^r$ from the null condition, $k^\mu k_\mu = 0$. Once again, observers are placed at present time, with random spatial positions inside random structures.
\newline\indent
The initial conditions for $k^\mu_{,\nu}$ are set according to the following procedure, requiring $\kappa^\mu = 0$ initially and ensuring that the null-condition and its partial derivative, as well as the definitions $dk^\mu/d\lambda := k^\nu k^\mu_{;\nu}$ are fulfilled. This ensures that our initial conditions are similar to those of \cite{selv_dzLTB}, justifying the assumption that our results correspond to a comoving emitter. Even when employing all the constraints, we are left with several $k^\mu_{,\nu}$ that can be chosen freely. Note that only $k^\mu_{,t}$ are important for $\delta z$ while the remaining $k^\mu_{,\nu}$ are gauge choices.
\newline\indent
To simplify calculations, we start by setting\footnote{We here make a specific choice for $k^\phi_{,t}$ which is one of the partial derivatives that can impact $\delta z$. For our specific setup, the choice $k^\phi_{,t} = 0$ is self-consistent and in line with the assumption of a comoving emitter.}
\begin{align}
    k^\phi_{,\mu} = k^\mu_{,\phi} = 0
\end{align}
along with
\begin{align}
    k^t_{,r} = 0.
\end{align}
Next, in Cartesian coordinates we set $k^i_{,j} = 0$, and use a coordinate transformation to obtain the corresponding  $k^\theta_{,r}$ as
\begin{align}
\begin{split}
    k^\theta_{,r} &= \frac{\dot{a}}{r^2 a} \big( \cos{\left( \theta \right)} \cos{\left( \phi \right)} x k^x \\ &+ \cos{\left( \theta \right)} \sin{\left( \phi \right)} y k^y - \sin{\left( \theta \right)} z k^z \big)- \frac{k^\theta}{r}.
\end{split}
\end{align}
We choose this initial condition for $k^\theta_{,r}$ even though the initial conditions are set in the LTB region. We may do this since we are here merely setting one of the free $k^\mu_{,\nu}$. Note that we will from now on {\em not} require $k^i_{,j} = 0$. We merely used this to motivate a choice for $k^\theta_{,r}$.
\newline\indent
We now implement constraints. We first require $\kappa^\mu = 0$ at the position of the observer. From earlier initial conditions, $\kappa^\phi = 0$ is already fulfilled, and $\kappa^t =0$ is trivially fulfilled. The requirement $\kappa^\mu = 0$ furthermore leads to the initial conditions
\begin{align}
    k^r_{,t} & = - \Gamma^r_{tr} k^r \\
    k^\theta_{,t} & = - \Gamma^\theta_{t\theta} k^\theta.
\end{align}
Using the total derivative constraints ($dk^\mu/d\lambda = k^\nu k^\mu_{;\nu}$), we can now obtain
\begin{align}
    & k^t_{,\theta} = \frac{1}{k^\theta} \left( \frac{dk^t}{d \lambda} - k^t k^t_{,t} - k^r k^t_{,r} \right) \\
    & k^\theta_{,\theta} = \frac{1}{k^\theta} \left( \frac{dk^\theta}{d \lambda} - k^t k^\theta_{,t} - k^r k^\theta_{,r} \right) .
\end{align}
From the partial derivatives of the null condition, we then find
\begin{align}
    & k^t_{,t} = \frac{1}{2 c^2 k^t} \left( A_t (k^r)^2 + B_t (k^\theta)^2 + 2 \left( A k^r k^r_{,t} + B k^\theta k^\theta_{,t} \right) \right) \\
    & k^r_{,r} = - \frac{1}{2 A k^r} \left( A_r (k^r)^2 + B_r (k^\theta)^2 + 2 \left( - c^2 k^t k^t_{,r} + B k^\theta k^\theta_{,r} \right) \right) \\
    & k^r_{,\theta} = - \frac{1}{A k^r} \left( - c^2 k^t k^t_{,\theta} + B k^\theta k^\theta_{,\theta} \right).
\end{align}
Initial conditions for all $k^\mu_{,\nu}$ are now given, and we confirm numerically that the full set fulfills the earlier-mentioned constraints.
\newline\newline
Except for the initial requirement $\kappa^\mu = 0$, the conditions we used to set initial conditions are constraints of our system of ODEs, and we track them when solving the ODEs in order to check the precision of our results. Around a handful of light rays broke the conditions to order $10^{-1}$ without any apparent explanation. We removed these rays from our samples and added new (random) rays with high precision fulfilment of the constraints.
\newline\newline
When propagating light rays through the Swiss-cheese model, we transform between a global coordinate system of the box and local coordinates of the LTB models. The details of these coordinate transformations are given in appendix \ref{app:coordtrans}.

\section{Numerical results}\label{sec:results}
This section serves to present the results from propagating light rays through our Swiss-cheese model and calculating the redshift drift along the light rays. We will first present results from individual fiducial light rays with the observer placed in either the cheese (FLRW background) or the hole (LTB structure). We will then move on to show the statistical results obtained by considering 100 light rays with the observer placed randomly in the cheese versus randomly in holes.
\newline\newline
Figure \ref{fig:single_ray} shows the integral over the individual terms of the integral in equation \ref{eq:integral_dz} for two fiducial observers. The figure clearly demonstrates how the deviation of the monopole term $\Sigma^{\bf O}$ from its EdS value almost exactly cancels with the quadrupole term $e^\mu   e^\nu \Sigma^{\bf ee}_{\mu \nu}$. These two terms will, in the following, be referred to as the Ricci and Weyl terms, respectively, since they are proportional to the Ricci and Weyl tensor (respectively). There is no clear cancellation between the remaining two terms, which are proportional to the shear and contracted position drift. These contributions are, however, subdominant by a factor of approximately 400 for the light ray with the observer placed in the cheese. For the observer placed in a hole, they are subdominant ``only'' by a factor of about 100. Although noticeable, this difference (400 versus 100) has no consequences since in both cases, the terms are clearly subdominant.
\newline\indent
\begin{figure*}[tb]
    \centering
    \includegraphics[width=0.48\textwidth]{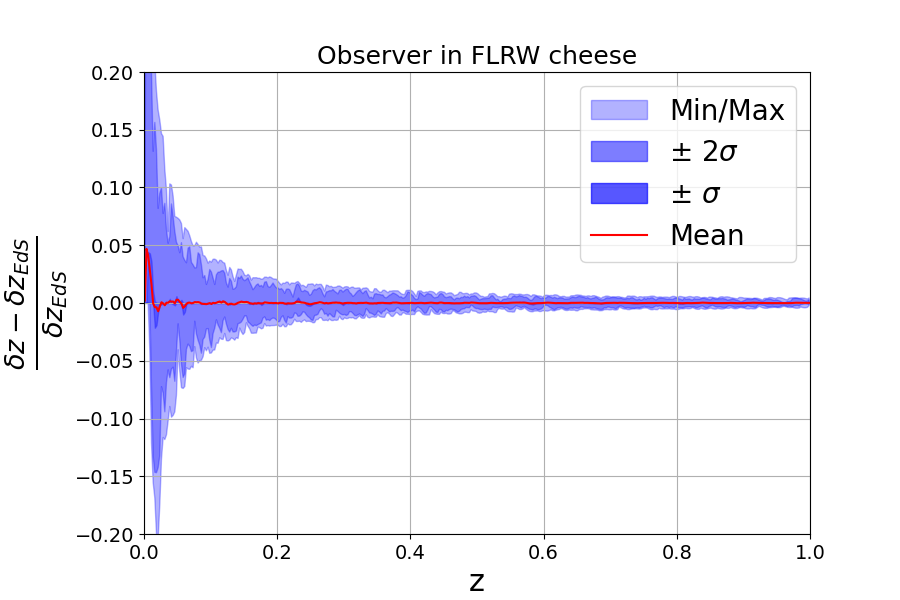}
        \includegraphics[width=0.48\textwidth]{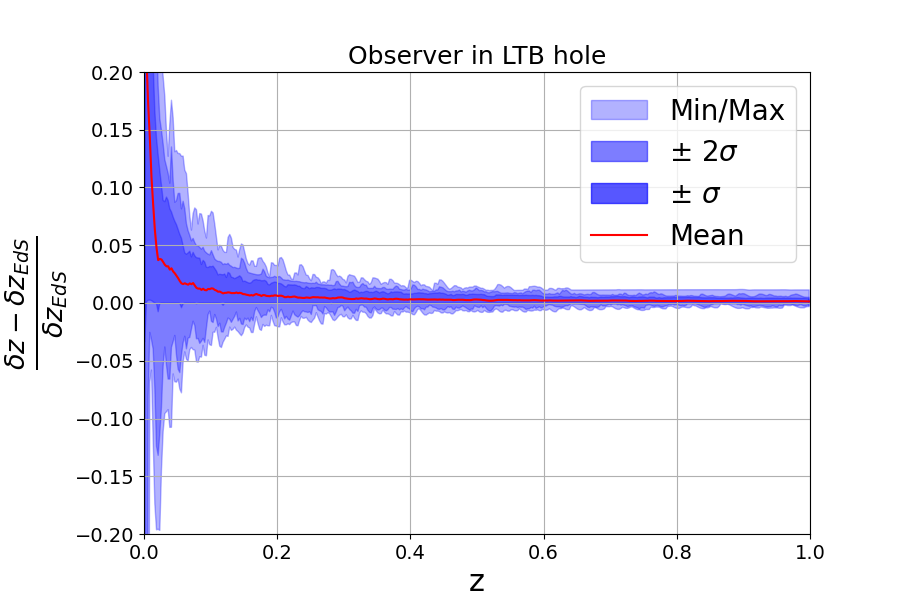}
    \caption{Mean and standard deviation of the redshift drift along 100 light rays with the observers placed in the cheese and the LTB structures, respectively.}
    \label{fig:100_rays}
\end{figure*}
The cancellation of the dominant fluctuations means that the redshift drift is everywhere close to that of the background EdS model. This is clearly seen in the bottom panels of figure \ref{fig:single_ray}, which show the relative deviation between the redshift drift of the Swiss-cheese model and the EdS model. Although the local fluctuation in the redshift drift is quite large ($\sim 10\%$) when the rays propagate through the first few structures, the fluctuations quickly become much smaller.
\newline\newline
Establishing the shear and position drift terms as subdominant is in line with earlier results based on single LTB models but to ensure that this result is not due to the fiducial rays being special, we will in the following section compare the terms along 100 light rays for two types of observations: One set of 100 observations with present-time observers placed randomly in the cheese and one set of 100 present-time observers placed randomly in holes.

\subsection{Multiple light rays}
Figure \ref{fig:100_rays} shows the mean redshift drift along 100 light rays with random present-time observers in the FLRW cheese and 100 light rays with observers placed in the LTB structures. For the observer placed in the cheese, the mean is very close to zero except at very low redshifts, where local effects dominate the signal. The fluctuations around the mean are somewhat large, still being a few percent at $z = 0.2$ when considering 2$\sigma$ around the mean\footnote{In all cases, the 1$\sigma$ and 2$\sigma$ intervals are approximated by the 16th–84th and 2.5th–97.5th percentiles of the distributions, respectively.}. However, the 1$\sigma$ interval around the mean is very narrow and almost indistinguishable from the mean. The fluctuations are noticeably larger when the observers are placed in the LTB structures, and in this case, the fluctuations are still near percent-level at $z = 1$, and the 1$\sigma$ band is now clearly visible. If we only consider the max/min and 2$\sigma$ intervals, the difference between the results obtained with the two types of observers is small.
\newline\indent
The mean visibly deviates from 0 until a much higher redshift when the observers are placed in holes rather than when they are placed in the cheese. We note that the mean for the LTB-based observers is slightly positive. This is the opposite of what was found in \cite{selv_alex}, where a small negative offset in the mean redshift drift was identified. That offset was conjectured to possibly be an ISW-type effect. Such an effect would, at the linear level, not be present here since we have not included a cosmological constant in our model. We cannot rule out that the offset is due to a nonlinear Rees-Sciama-type effect, but it seems more likely that the offset we find is due to local effects. Indeed, unless the observer is placed close to the symmetry center of an LTB structure, there will locally be a significant anisotropy, which can lead to a large Weyl contribution, which can again lead to the offset. A large Weyl contribution is e.g. seen in figure \ref{fig:single_ray} for the observer placed in an LTB hole. Regardless its reason, the deviation of the mean from zero is small. Specifically, at $z = 1$, the means are $\sim 0.001$ and $\sim 0.004$, for observers in the FLRW cheese and LTB holes, respectively.
\newline\indent

\begin{figure*}[tb]
\centering
\includegraphics[scale = 0.38]{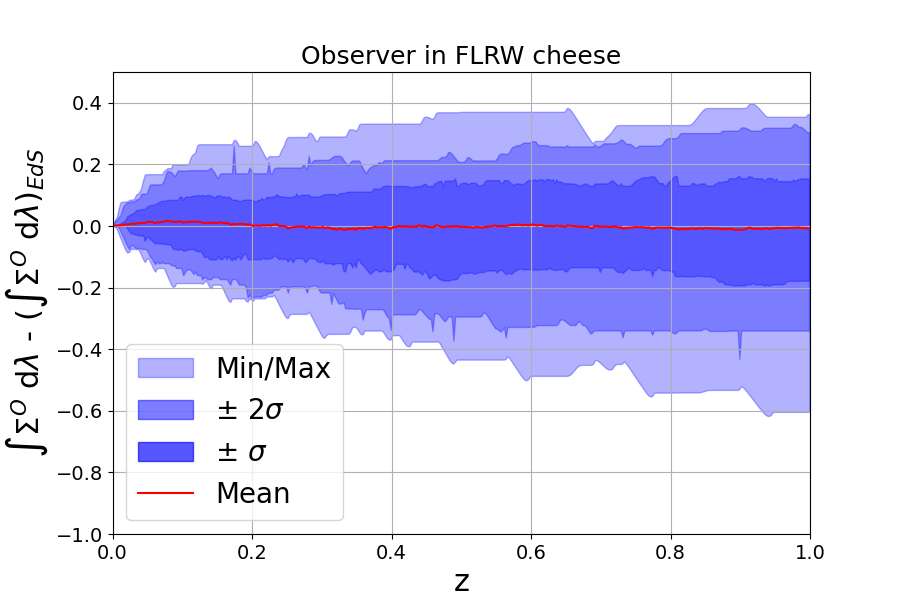}
\includegraphics[scale = 0.38]{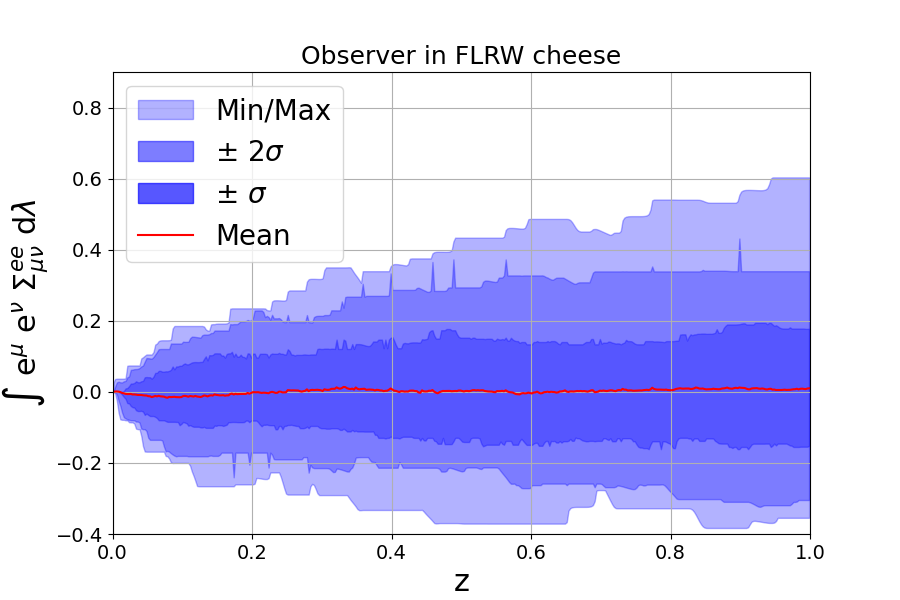}\\
\includegraphics[scale = 0.38]{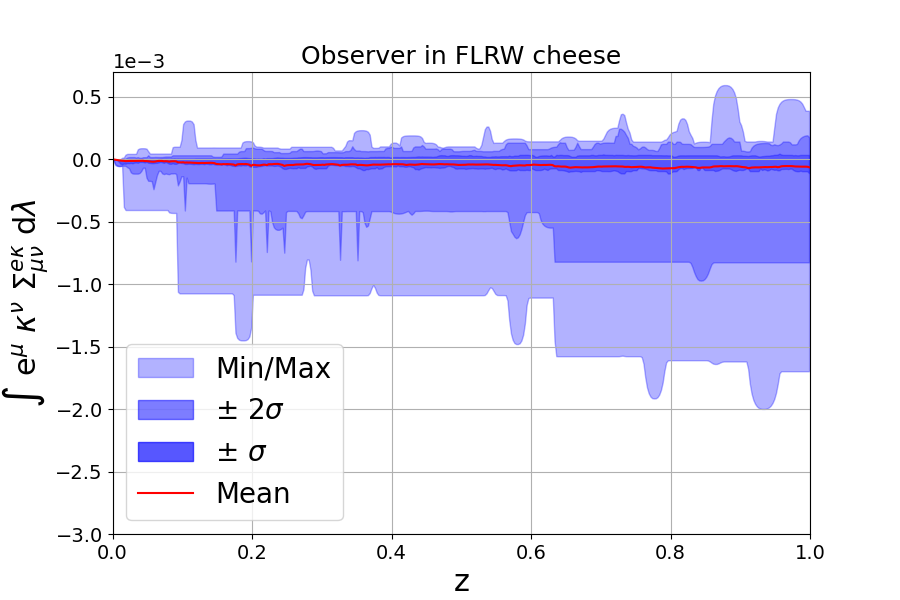}
\includegraphics[scale = 0.38]{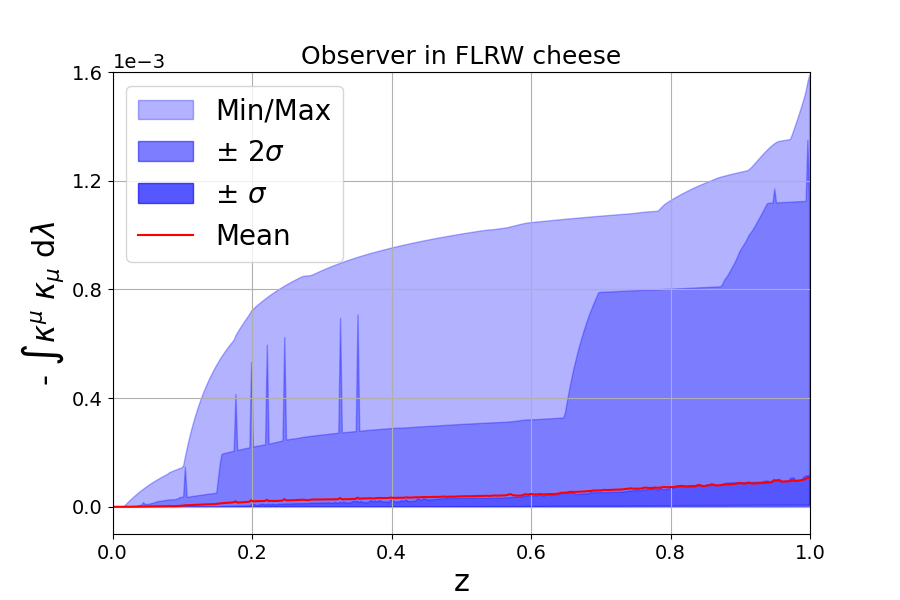}
\caption{Mean and standard deviation of the individual contributions to the redshift drift (equation \ref{eq:integral_dz}) along 100 light rays with the observers placed in the FLRW cheese.}
\label{fig:100_components_cheese}
\end{figure*}

\begin{figure*}[tb]
\centering
\includegraphics[scale = 0.38]{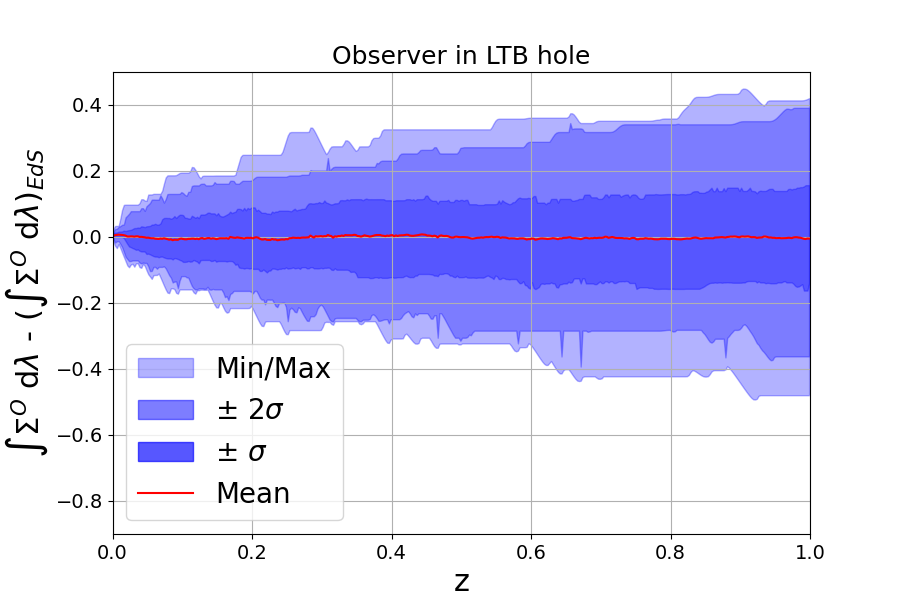}
\includegraphics[scale = 0.38]{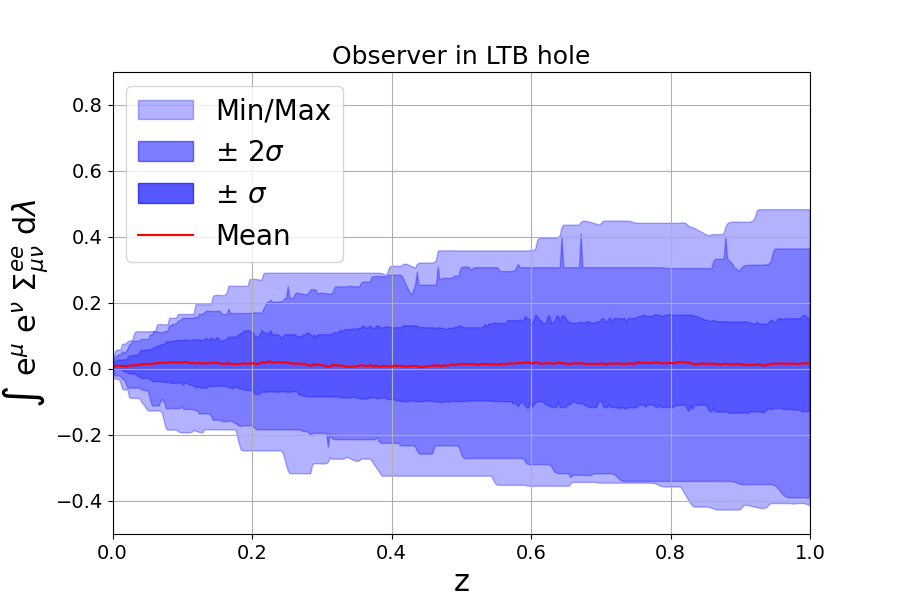}\\
\includegraphics[scale = 0.38]{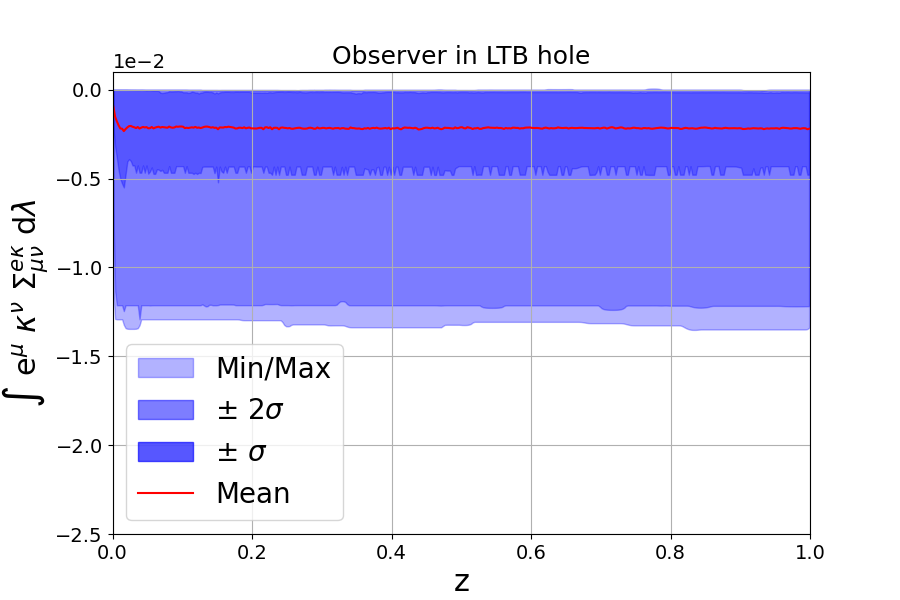}
\includegraphics[scale = 0.38]{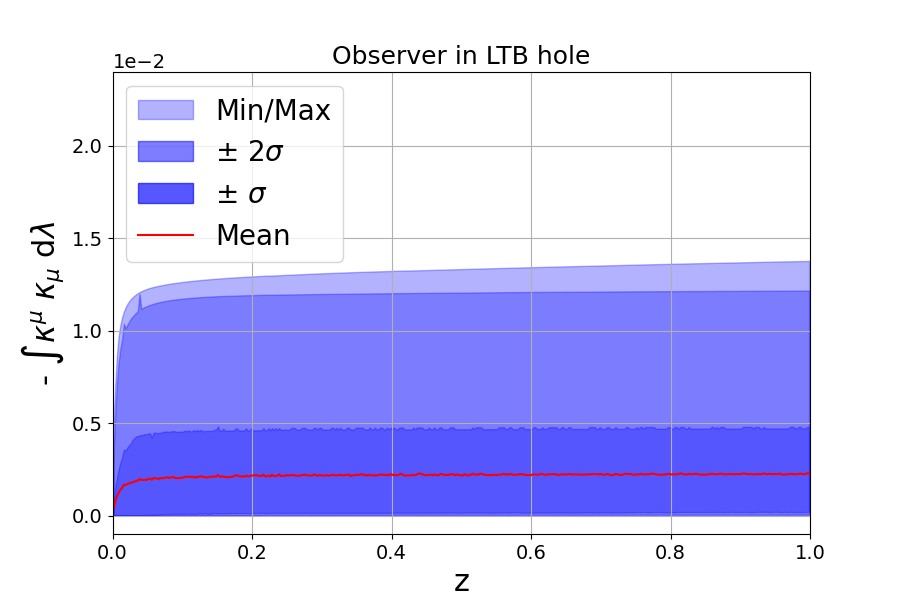}
\caption{Mean and standard deviation of the individual contributions to the redshift drift (equation \ref{eq:integral_dz}) along 100 light rays with the observers placed in the LTB structures.}
\label{fig:100_components_hole}
\end{figure*}

Figure \ref{fig:100_components_cheese} shows the mean and fluctuations of the four contributions to the redshift drift for 100 light rays with cheese-based observers. The means of the two dominant components (the Weyl contribution and the Ricci contribution's fluctuations around the background value) nearly vanish and would anyway cancel each other, consistent with our findings from considering individual rays. The two subdominant components both have means that clearly deviate from zero, though only at order $10^{-4}-10^{-3}$. Thus, although these two terms in principle lead to small changes in the redshift drift, even their fluctuations around their mean are consistently so small that they would only be important for very high-precision measurements.
\newline\indent
Note also that the position drift term has a mean that is dominated by outliers, which leads the mean to lie slightly outside the 1$\sigma$ interval.
\newline\indent
Figure \ref{fig:100_components_hole} shows the mean and fluctuations of the four contributions to the redshift drift integral for 100 light rays with observers placed in the LTB structures. The dominant (Ricci and Weyl) contributions to the redshift drift are here very similar to the case with cheese-based observers. However, the fluctuations in the shear and position drift terms have increased by roughly an order of magnitude, and are now about 1\% the size of the fluctuations in the Ricci and Weyl terms. Their means also deviate visibly from zero but are around the same size as before. In this case, the term proportional to the shear appears to be almost entirely non-positive. This was not the case along light rays observed from the cheese. We expect this difference to be simply due to large negative shear contributions being dominant for observers in the holes.
\newline\indent
In all cases, the Ricci and Weyl terms are clearly dominant, and although the subdominant contributions in principle lead to offsets in the total redshift drift, these terms are orders of magnitude smaller than the dominant contributions.

\section{Summary, discussion and conclusions}\label{sec:summary}
We constructed a Swiss-cheese cosmological model and computed the redshift drift along 200 light rays, with observers placed either in the homogeneous FLRW regions (100 light rays) or within the inhomogeneous LTB structures (100 light rays). For each configuration, we evaluated the mean and percentile distributions of the total redshift drift and its different components.
\newline\indent
Across all cases, we find the two dominant contributions to the redshift drift fluctuations (those associated with Ricci and Weyl components) nearly cancel with each other, resulting in a total redshift drift that closely matches the Einstein-de Sitter (EdS) background prediction. Significant deviations occur only at very low redshifts, where light rays traverse structures near the observer, leading to fluctuations of several tens of percent. The subdominant contributions, arising from shear and position drift, are an order of magnitude larger when observers are located within LTB structures instead of in the FLRW region. Nevertheless, these contributions remain below or around 1\% of the total redshift drift signal. While their means do not vanish, the resulting offsets relative to the FLRW background are well below the percent level, making it unlikely that they would bias measurements from upcoming facilities such as SKA or ELT.
\newline\indent
Previous studies \cite{selv_ET, selv_gadget} assumed the dominant contributions to redshift drift arise from Ricci and Weyl terms, an assumption supported by the single-structure analysis in \cite{selv_dzLTB}. However, this earlier work did not assess whether subdominant terms accumulate over extended redshift intervals since only individual LTB structures were considered. Our results show that such accumulation does occur, but remains slow: even at redshift $z = 1$, the redshift drift can be accurately approximated using only the dominant Ricci and Weyl contributions to good precision.
\newline\indent
Although our analysis is based on an idealized Swiss-cheese model with spherically symmetric LTB structures, we expect the qualitative results to extend to more realistic scenarios, such as cosmological N-body simulations, provided appropriate precautions are taken. Most importantly, many terms in the expression for the redshift drift in equation \ref{eq:integral_dz} vanish identically for LTB models but may be important in more realistic settings. For instance, we would expect that peculiar acceleration can have a significant impact on the redshift drift as found in e.g. \cite{selv_alex}. However, most importantly, we expect that vorticity remains small above the scales of virialization. Therefore, we expect that terms involving $\kappa^{\mu}$ and hence the partial derivatives of the tangent vector, $k^{\mu}_{,\nu}$, can still be neglected in e.g. realistic N-body and hydrodynamical simulations.
\newline\indent
We do not expect the near-cancellation of Ricci and Weyl terms to be an artifact of spherical symmetry, as similar behavior has been observed for other observables such as redshift and angular diameter distance in models lacking spherical symmetry \cite{selv_packing}. Lastly, neglecting a cosmological constant in our model has no impact on our findings since the addition of such a homogeneous contribution would only add to the background part of the redshift drift signal.

\begin{acknowledgments}
We thank Alexander Oestreicher for comments on the manuscript. This project was funded by VILLUM FONDEN, grant VIL53032 (PI: SMK).
\newline\newline
{\bf Author contribution statement}: The presented work builds upon the results of the master’s thesis project of DRS, conducted under the supervision of SMK. DRS carried out all theoretical and numerical calculations under the guidance of SMK, including occasional comparisons with independent results obtained by SMK. Both authors contributed significantly to the writing of the manuscript.
\end{acknowledgments}

\appendix

\section{Random close packing algorithm}\label{app:packing}
In this appendix, we provide the details of the algorithm we used for distributing LTB structures in our Swiss-cheese model. Our algorithm follows the Jodrey-Tory algorithm of \cite{torey1, torey2}.
\newline\newline
We first randomly set the positions of the centers of $N$ spheres. We then define an {\em outer diameter}
\begin{align}
     d_{out}^0 = L \left( \frac{6 \chi}{\pi N} \right)^{\frac{1}{3}} ,
\end{align}
where $L$ is the length of the box (assumed to be equal along all axes), and $\chi$ is a free parameter representing the packing fraction, that is, the ratio of the total volume of the spheres to the volume of the box. We will denote the smallest distance between all center pairs as the inner diameter, $d_{\rm in}$. The outer diameter, $d_{out}^0$, is the target packing diameter, which is the diameter each sphere must have for the packing to be complete. To achieve this, we repeat the following steps until the inner diameter is greater than the outer diameter.
\begin{itemize}
    \item Find the smallest distance between centre pairs and set it equal to the inner diameter, $d_{\rm in}^i$. Check whether $d_{\rm in}>d_{\rm out}$ and if so, stop. Otherwise, continue to the next step.
    \item Move the closest pair away from each other in the direction of the line joining the two.
    \item Recalculate the shortest distance from the two new center positions to all other centers.
    \item Update the outer diameter using

\begin{align}
    d_{\rm out}^{i+1} = d_{\rm out}^i - \left( \frac{1}{2} \right)^{\delta} \frac{d_{\rm out}^0}{\gamma} \label{new_outer} \; ,
\end{align}
where $\gamma$ is the decay rate of $d_{\rm out}$ and 
\begin{align}
    \delta = \text{floor}\left( - \log_{10}{\left( \Delta \chi \right)} \right) \; ,
\end{align}
with $\Delta \chi$ being the difference in the packing fraction between $d_{\rm out}$ and $d_{\rm in}$
\begin{align}
    \Delta \chi = \frac{\pi N}{6 L^3} \left( \left( d_{out}^i \right)^3 - \left( d_{in}^i \right)^3 \right) \; .
\end{align}
\end{itemize}
The algorithm is demonstrated in Figure \ref{fig:packing_algorithm}.

\section{Coordinate transformations}\label{app:coordtrans}
In this appendix, we provide the details of moving a light ray between a local and global coordinate system in the Swiss-cheese model, along with the equations needed to move between Cartesian and spherical coordinates.
\newline\newline
When a light ray reaches the boundary between the EdS cheese and an LTB model, we need to switch between the global coordinate system of the Swiss-cheese model and the local coordinate system of the LTB model. This simply requires translating the Cartesian coordinates. Since the global coordinate system is Cartesian while the local is spherical, we also need to switch between these two types of coordinates.
We use the standard convention
\begin{align}
    & x = r \sin{\left( \theta \right)} \cos{\left( \phi \right)} \\
    & y = r \sin{\left( \theta \right)} \sin{\left( \phi \right)} \\
    & z = r \cos{\left( \theta \right)} \\
    & r = \sqrt{x^2 + y^2 + z^2} \\
    & \theta = \text{atan2}\left( \sqrt{x^2 + y^2}, z \right) \\
    & \phi = \text{atan2}\left( y, x \right) \\
\end{align}
For transforming $k^\mu$ we use the usual tensor transformation rule, e.g. $k^{\mu, car} = \frac{d x^{\mu, car}}{d x^{\lambda, sph}} k^{\lambda, sph}$. The partial derivative of the tangent vectors cannot be directly transformed. However, we can switch between partial and covariant derivatives using the definition
\begin{align}
    \nabla_\sigma k^\mu := k^\mu_{,\sigma} + \Gamma^\mu_{\sigma \nu} k^\nu.
\end{align}
The covariant derivative is then transformed according to the usual tensor transformation rule
\begin{align}
    \nabla_\nu k^{\mu, car} = \frac{d x^{\mu, car}}{d x^{\lambda, sph}} \frac{d x^{\kappa, sph}}{d x^{\nu, car}} \nabla_{\kappa} k^{\lambda, sph}.
\end{align}
Afterwards, we again use the relation between partial and covariant derivatives to obtain the transformed partial derivatives.

\clearpage

\begin{center}
\begin{minipage}{0.9\textwidth}
    \centering
    \includegraphics[width=0.8\textwidth]{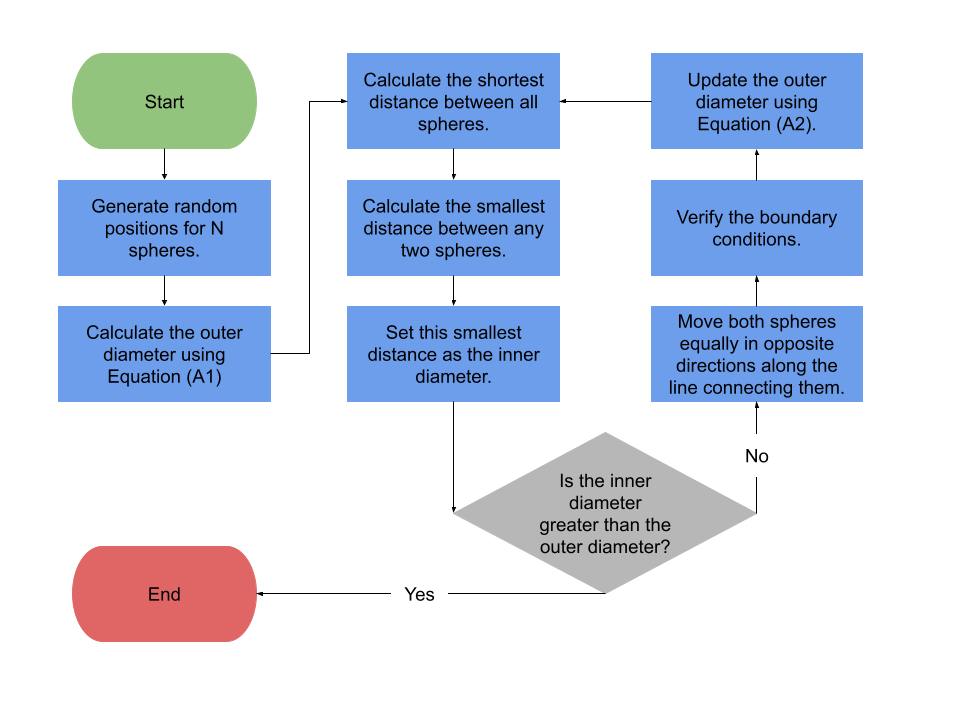}
    \captionof{figure}{Flow chart demonstrating the algorithm for distributing LTB structures in the EdS cheese.}
    \label{fig:packing_algorithm}
\end{minipage}
\end{center}

\end{document}